\documentclass[10pt,prd,tightenlines,twocolumn,superscriptaddress]{revtex4}

\usepackage{amsmath,amssymb}

\usepackage{color}
\usepackage{graphicx}
\definecolor{darkblue}{rgb}{0,0,0.7}
\definecolor{darkred}{rgb}{0.7,0,0}
\definecolor{darkgreen}{rgb}{0,0.5,0}
\definecolor{added}{rgb}{0.8,0.4,0} 
\usepackage[unicode, colorlinks, citecolor=darkblue, linkcolor=darkred, urlcolor=blue]{hyperref}

\begin{document}

\title{Squeezing of optomechanical modes in detuned Fabry-Perot interferometer}

\author{Andrey A. Rakhubovsky \footnote{Electronic address: rkh@hbar.phys.msu.ru}}
\affiliation{Physics Department, Moscow State University, Moscow 119992 Russia}

\author{Sergey P. Vyatchanin}
\affiliation{Physics Department, Moscow State University, Moscow 119992 Russia}

\date{\today}

\begin{abstract}

	We carry out analysis of optomechanical system formed by movable mirror of Fabry-Perot cavity pumped by detuned laser. Optical spring arising from detuned pump creates in the system several eigen modes which could be treated as high-Q oscillators. Modulation of laser power results in parametric modulation of oscillators spring constants thus allowing to squeeze noise in quadratures of the modes. Evidence of the squeezing could be found in the light reflected from the cavity. 

\end{abstract}

\maketitle

\section{Introduction}

The purpose of gravitational waves detection leads to construction of large-scale antennas like LIGO~\cite{Abbott_2009,Harry_2010}, VIRGO~\cite{Accadia2012} and GEO~\cite{Grote2010}. Very high sensitivity of these devices is limited by a number of noises.  In the low frequency range ( below $\sim 50$ Hz) the prevailing sources of noise are seismic ones, at middle frequencies ($\sim 50 - 200$ Hz) thermal noises dominate and in high frequency range (over $200$ Hz) photon shot noise prevails. However the technical improvement of antennas by compensation and suppression of these and other noises will allow to achieve sensitivity level defined only by quantum noise which for continuous position measurement has lowest boundary defined by Standard Quantum Limit (SQL) \cite{1968_SQL,1975_SQL,1977_SQL,1992_quant_meas}. SQL is the optimal combination of two noises of quantum nature: fluctuations of mirror motion caused by random photon number falling onto its surface and photon counting noise.

One of the ways to overcome the SQL is the implementation of so-called optical rigidity (optical spring) effect~\cite{64a1BrMiVMU,67a1BrMaJETP,78BrBook,1992_quant_meas} which is based on a fact that in a detuned Fabry-Perot interferometer the circulating power and consequently the radiation pressure is dependent on the distance between the mirrors. It has been shown in a number of papers~\cite{99a1BrKhPLA,01a1KhPLA,01ChenPRD,02ChenPRD,05a1LaVyPLA,06a1KhLaVyPRD,2007_NiSaKa_PRD} that interferometers using optical springs exhibit sensitivity below the SQL.

In a system utilizing optical spring there are two degrees of freedom (in case of one pump): a mechanical one and an optical one. Interaction of these coordinates gives birth to several eigen modes (the number of which is equal to the system degrees of freedom number) each of which is characterized by its own resonance frequency and damping.   

Free evolution of the system can be represented as a sum of eigen modes each of which is an oscillator with its eigen frequency and has corresponding damping. In principle one can make transfer from the conventional coordinates to eigen ones and consider the evolution of the system as evolution of these new oscillators. 

Using the curve of susceptibility of the system one can estimate the average energy stored in each equivalent eigen mode oscillator. It is well known~\cite{1968_SQL,1992_quant_meas,1996_QND_toys_tools} that an oscillator with average energy $E_\text a$ exhibits quantum-mechanical behavior if the condition 
\begin{equation}
	\notag
	\frac{E_\text{a} }{ \hbar \omega Q } \ll 1.
\end{equation}
is met. Here $\omega$ and $Q$ are eigen frequency and quality factor of oscillator.  Estimations~\cite{2012_RaVy_PLA} show that for system such as interferometer Advanced LIGO these conditions are well fulfilled. As a consequence one can expect the corresponding eigen modes to be observed in quantum state despite the fact that the (effective) mass of mechanical oscillator in considered interferometer is equal to 10 kg.  

Given this motivation it is interesting to look for any experimental scenario giving an ability to bring one of the eigen modes to non-classical state. Currently the development of techniques aimed on preparation of mechanical resonator in quantum state by means of optomechanical interaction is well underway including investigations of micromembranes~\cite{Thompson2008}, microtoroids~\cite{Verhagen2012}, optomechanical crystals~\cite{Eichenfield2009}, pulse-pumped optomechanical cavities~\cite{Vanner2011} and even large-scale gravitational-wave detectors~\cite{Khalili2010}. One of the most important problems inherent to optomechanical devices is relatively large losses of mechanical microoscillator. Replacement of material spring by optical one may decrease the losses in mechanical system and thus provide an experimental possibility of realisation of quantum regimes.

For the initial consideration it is also worth to show the very possibility of manipulation with optomechanical modes. To prove this possibility we demonstrate the mechanism to perform quadrature squeezing of shot noise caused fluctuations in these modes using parametric modulation of spring constant which has been previously considered as a tool to squeeze fluctuations in cavity mirror motion~\cite{Szorkovszky2012}.

\section{Description of model}

\begin{figure}[ht]
	\begin{center}
		\fbox{ \includegraphics[width = .95 \linewidth]{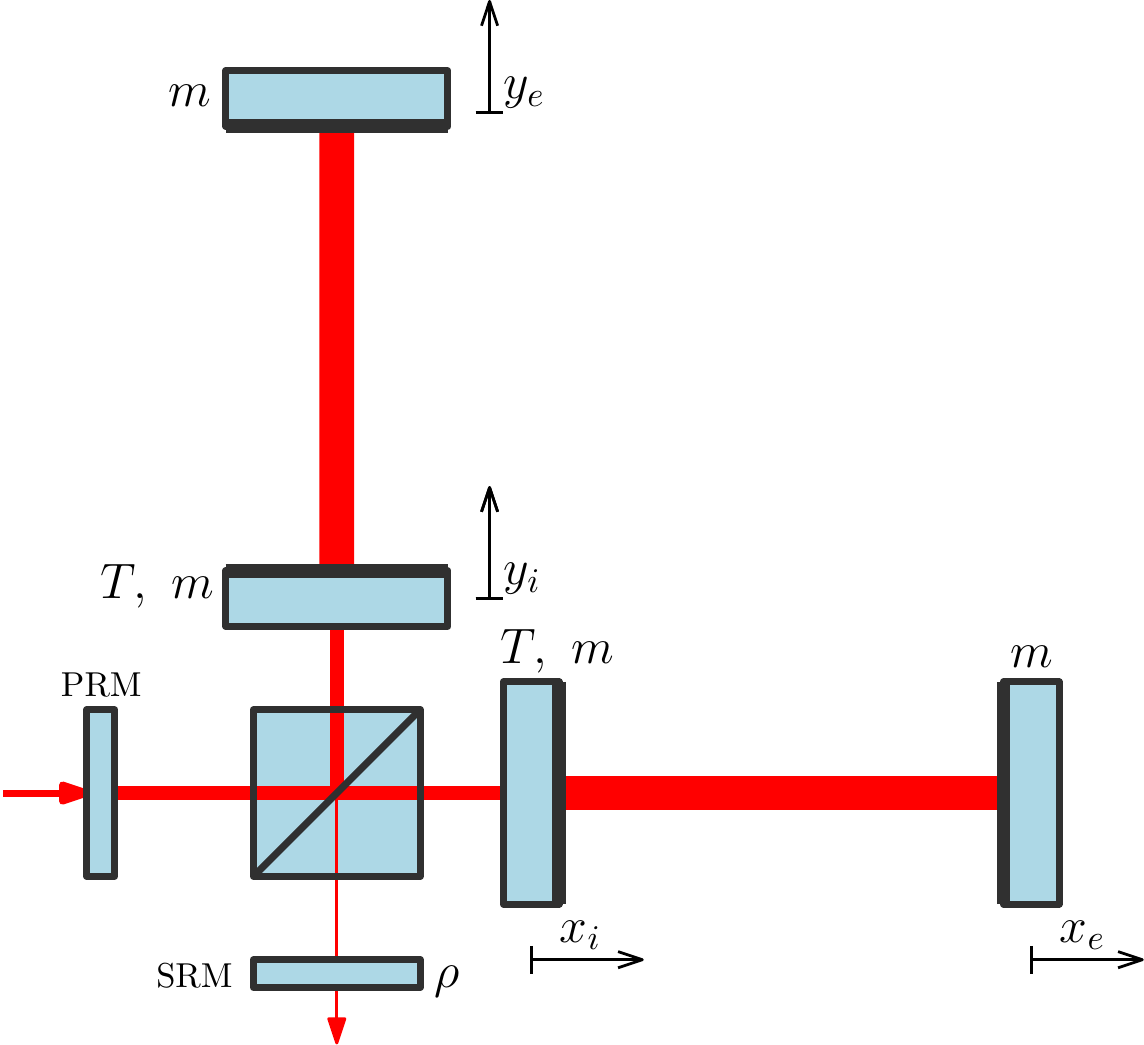}}
		\fbox{ \includegraphics[width = .95 \linewidth]{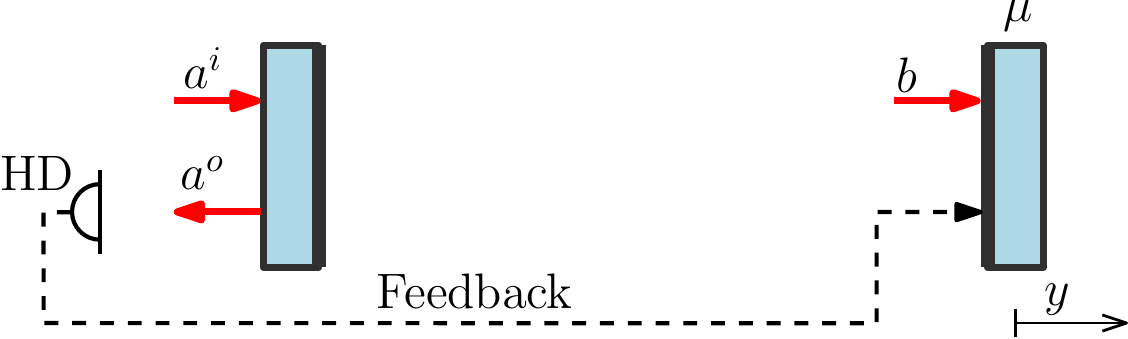}}
		\caption{
		Top: scheme of Advanced LIGO detector. Bottom: scheme of equivalent Fabry-Perot interferometer with one movable mirror and feedback implemented for stability of system. 
		}
		\label{fig:aLIGO}
	\end{center}
\end{figure}

We consider a gravitational-wave detector Advanced LIGO with a signal recycling mirror (SRM) having an amplitude reflectivity $\rho$ and power recycling mirror (PRM); a scheme of antenna is presented in fig.~\ref{fig:aLIGO}. Antenna consists of a Michelson interferometer with additional mirrors forming Fabry-Perot (FP) cavities with mean distance $L$ between mirrors in arms. Input mirrors have amplitude transmittivities $T$ and masses $m$ and output mirrors have the same masses $m$ and are completely reflective. We assume that all mirrors are lossless. The interferometer is pumped by laser having frequency resonant for FP cavities in arms.

Dynamics of this interferometer could be split into two modes: namely differential one and common one. The optical fields in the modes represent difference and sum of the fields in arms respectively and carry information about difference and sum between lengthenings of arm cavities. Each mode is characterized by optical detuning and decay rate introduced by displacement of corresponding recycling mirror: SRM for differential mode and PRM for common one. 

It has been demonstrated~\cite{02ChenPRD} that each of this modes could be
equivalently described using a model of FP cavity (see fig.~\ref{fig:aLIGO},
bottom). We focus our attention to differential mode.  The end mirror of
equivalent cavity has mass equal to reduced mass of four mirrors of Michelson
interferometer \smash{$\mu = \frac{m}{4}$} and is completely reflective. The
displacement of this mirror is equal to difference of arms lengthening of
initial interferometer: $y = (x_e - x_i) - (y_e - y_i)$. Distance separating
the mirrors is equal to one in initial interferometer, namely $L$. The pump in
equivalent scheme should be characterized by the same relaxation rate $\Gamma$
and detuning $\Delta$ as in initial scheme and mean circulating power $P$ two
times bigger than one in initial interferometer. 

A detuned pump creates in FP cavity an optical spring, that is the
radiation-pressure force which depends on the distance separating mirrors of
the cavity. A system with only one optical spring is always unstable because a
single pump introduces either negative damping or negative
rigidity~\cite{64a1BrMiVMU, 67a1BrMaJETP, 78BrBook,99a1BrKhPLA}. A few ways to
avoid instabilities have been proposed amongst which there are implementation
of feedback~\cite{02ChenPRD} or utilization of additional pump
\cite{08ChenPRD,2011_RaHiVy_PRD}. The latter way has been investigated in
details and proven experimentally with mirror of gram-scale~\cite{07CorbitPRL}.
However in laboratory-scale experiment it should be easier to utilize a proper
feedback. Hence we consider a scheme of FP cavity with one pump and a feedback
instead of two pumps. 

Input optical fluctuations are described by annihilation operator~$a^i$,
operator $a^o$ describes output fluctuations. To avoid instability caused by
optical spring we measure phase quadrature $a_2^o$ of output fluctuations with
homodyne detector and send its derivative over time as a feedback force to the
movable mirror.

The dynamics of interferometer could be described in terms of two degrees of freedom. One of those is annihilation operator $b$ of optical fluctuations inside the interferometer (actually, its amplitude quadrature $b_1$) and another is mechanical displacement $y$. Equations of motion for this system could be written as follows (in capacity of mechanical coordinate we use dimensionless displacement $z$):
\begin{subequations}
	\label{eq:sys1}
	\begin{align}
		& \ddot b_1 + g \dot b_1 + 2 b_1 + A z  = \nu_1;
		\\
		& \ddot z - A b_1 - \alpha \dot  b_1  = \nu_2. 
	\end{align}
\end{subequations}
{
Derivation of these equations from the hamiltonian of the system and exact definition of coefficients are presented in Appendix~\ref{app:notations}. }

We use dimensionless parameters defined the same way as
in~\cite{2011_RaHiVy_PRD,2012_RaVy_PLA}. Parameter $g$ stands for optical
relaxation rate, coupling coefficient $A$ is proportional to pump power,
$\alpha$ is the coefficient of feedback. The right parts of equations represent
fluctuational forces acting on corresponding degree of freedom. As of optical
one the force $\nu_1$ represents input vacuum fluctuations and for mechanical
one $\nu_2$ describes vacuum fluctuations reflected by the cavity and
transmitted by feedback. We neglect all other fluctuational forces including
thermal and seismic ones. 
{Depending on device in consideration this first step assumption could be either realistic or not. Some calculations of thermal noise influence is presented in Appendix~\ref{app:thermal}. In particular they show that in configuration of Advanced LIGO interferometer the impact of thermal noises is only slightly below the one of quantum fluctuations. }


The equations~\eqref{eq:sys1} look similar to ones for oscillator coupled to a
free mass (or two coupled oscillators one of which has partial frequency equal
to zero). However we would note the difference in signs in front of the
coupling terms (proportional to $A$): in the case of conventional coupled
oscillators these signs coincide; in our case the signs differ. This is the
reason of instability which appears in absence of feedback and this is also the
reason of possibility for eigen frequencies of such system to coincide (so
called double resonance case \cite{01a1KhPLA, 05a1LaVyPLA,06a1KhLaVyPRD} ---
let us remind that in case of two coupled oscillators coincidence of eigen
frequencies is impossible). 
{
\section{Parametric squeezing of eigen modes amplitudes}
}
As in the case of conventional coupled oscillators one can treat the system in terms of eigen modes. The evolution of oscillator system could be written as a sum of eigen oscillations: 
\begin{equation}
	\label{eq:decomposition}
	\begin{pmatrix}
		b_1 \\ z 
	\end{pmatrix} = 
	\sum_i^2 \vec v_i g_i  e^{ - i \omega_i t } + \vec v_i^* g_i^\dagger  e^{i \omega_i t }. 
\end{equation}
Index in the sum runs from 1 to the number of system degrees of freedom (2 for set \eqref{eq:sys1}). This equation could be treated as transformation from conventional coordinates $(b_1; z)$ to the eigen ones. As in eqn.~\eqref{eq:decomposition} one can present evolution of each of eigen modes as combination of oscillations at corresponding eigen frequencies $\omega_i$ and slow (compared to these oscillations) changing of amplitude. In case of free evolution (zero right parts in eqns.~\eqref{eq:sys1}) quantities $g_i\sim e^{-\gamma_it}$ i.e. they freely decay  with relaxation rates $\gamma_i$ into the equilibrium values. Vectors $\vec v_i$ represent distribution of amplitudes in corresponding eigen modes. 

Suppose that we have a possibility to modulate the power of pumping laser. This will result in modulation of coupling coefficient and hence it will cause shifting of eigen frequencies of system. It is well known \cite{Collett1984} that modulation of eigen frequency of oscillator with frequency twice bigger than its own one results in squeezing of noise in its quadratures. In this paper we show that modulation of the pumping power performs squeezing of noise in quadratures of eigen modes amplitudes. 

Consider we harmonically modulate the pumping power at frequency $2p$ so coupling coefficient in~\eqref{eq:sys1} depends on time: $A \to A ( 1 + 2 |m| \cos ( 2 p t + \phi ))$. Assuming the quantities $g_i$ to be \emph{slow} (i.e. not to significantly change on times compared to mechanical periods \smash{$2 \pi/ \omega_i$}) one can plug the expression~\eqref{eq:decomposition} into system~\eqref{eq:sys1} and to get rid of rapidly oscillating terms by averaging over a period $2 \pi/ p $. This procedure results in shortened equations for amplitudes $g_i$. For one with number $j$ ($j = 1,2$) the equation takes form: 
\begin{multline}
	\label{eq:shorteq}
	- 2 i \omega_j \big[ \dot g_j + \gamma_j g_j \big] + |m| \sum_i g_i^\dagger ( \vec \Pi_j \vec w_i^* ) e^{  i ( \omega_i + \omega_j - 2 p ) t - i \phi } =
	\\
	= ( \vec \Pi_j {\underline{\nu}_j}) e^{i \omega_j t }. 
\end{multline}
Here we have introduced a set of vectors $\vec \Pi_j$ built to be orthogonal to all $v_i$ except one: $( \vec \Pi_j \vec v_i ) = \delta_{ji}$, where $\delta_{ji}$ is a Kronecker delta (in the case of two conventional coupled oscillators vectors $\vec v_i$ are itself orthogonal and system of $\vec \Pi_i$ is unnecessary). Vectors $\vec w_i$ are defined as follows:
\begin{equation}
	\notag
	\vec w_i = 
	\begin{pmatrix}
		0 & A \\ - A & 0 
	\end{pmatrix}
	\vec v_i. 
\end{equation}
Underlining used for $\underline{\nu}_j$ is to emphasise that this quantity is obtained in right part of equation by procedure of keeping only \emph{slow} (in comparison to terms oscillating with frequency $p$) quantities. To make $\nu_j e^{i \omega_j t }$ fulfilling this criterion one should keep in $\nu_j$ only spectral components close to $\omega_j$. 

From now on we will focus on the amplitude of one of modes taking into account that all the results obtained for this mode are similar for other ones. For clarity let us set $j=1$ thus considering the first mode.  Also we suppose the modulation to happen at frequency twice bigger than the one of first mode: $p = \omega_1$. 

First let us consider the simplest case when difference between eigen frequencies $\omega_1 - \omega_2$ is large in respect to decay rates $\gamma_i$. In this case the exponential multipliers in eqn.~\eqref{eq:shorteq} should be considered as fast oscillating ones and the equations for amplitudes $g_i$ decouple. In this case the equation for $g_1$ takes the following form in spectral domain 
\begin{equation}
	\label{eq:g1}
	\big[ \gamma_1 - ix ] g_1 (x) + \epsilon_{11} g_1^\dagger (-x) = \frac{ i ( \vec \Pi_1 \vec \nu_1) e^{ i \omega_1 t }}{ 2 \omega_1 } \equiv f_1 (x). 
\end{equation}
Here $x$ stands for normalized spectral frequency (see appendix~\ref{app:notations}), $\gamma_1$ is the decay rate of first mode; $\epsilon_{11}$ is a quantity proportional to modulation strength, the general expression for it could be easily deduced from eqn.~\eqref{eq:shorteq}: 
\begin{equation}
	\notag
	\epsilon_{ji} = \frac{- i |m| ( \vec \Pi_j^* \vec w_i ) e^{ i \phi }}{ 2 \omega_j }.
\end{equation}

We also write the equation for $g_1^\dagger ( - x)$ by hermitian conjugation of eqn.~\ref{eq:g1} and replacement $x \to -x$. Taking proper combinations of these equations yields the expressions for corresponding quadratures of $g_1$. For simplicity let $\epsilon_{11} = \epsilon_{11}^* = | \epsilon_{11}|$ which is achievable by proper choosing $\phi$, in this case the mentioned combinations reduce to sum or difference:

\begin{subequations}
	\begin{align}
		\label{eq:sumq}
		\big( g_1 + g_1^\dagger\big) \Big[ \gamma_1 + |\epsilon_{11} | - ix \Big] = f_1 + f_1^\dagger;
		\\
		\label{eq:difq}
		\big(g_1 - g_1^\dagger \big) \Big[ \gamma_1 - |\epsilon_{11} | - ix \Big] = f_1 - f_1^\dagger;
	\end{align}
\end{subequations}
{One can conclude} from these equations that sum quadrature is squeezed due to parametric modulation and the difference quadrature is antisqueezed. {The measure of squeezing is the spectral density which for arbitrary quantity $d(\omega)$ is given by expression 
\begin{equation}
	\notag
	S_d(\omega) 2 \pi \delta ( \omega + \omega') = 1/2 \cdot \langle d^\dagger(\omega) d (\omega') + d (\omega') d^\dagger(\omega')\rangle,
\end{equation}
where the angle brackets mean averaging.

Obvious calculations prove that spectral densities of right parts of both equations~\eqref{eq:sumq} and~\eqref{eq:difq} coincide and the equations differ by additional damping $| \epsilon_{11}|$  (positive or negative for different quadratures) introduced by modulation. In absence of modulation $\epsilon_{11} = 0$  and there is no discrepancy between quadratures. If modulation is applied then $\epsilon_{11}$ has nonzero value and quadratures spectral densities differ.}

One can also estimate the critical level of modulation characterized by value $m_\text{c}$, this is a level when negative damping added to the differential quadrature $g_1 - g_1^\dagger$ becomes equal to its own damping: $\gamma_1 -  | \epsilon_{11}| = 0$ thus making quadrature instable. Given this value of modulation coefficient damping in the sum quadrature is twice bigger than its own damping $\gamma_1$. This means that quadrature noise squeezing is limited by factor of two.

In general case of arbitrary eigen frequencies $\omega_{1,2}$ the equations~\eqref{eq:shorteq} could not be solved in such an obvious way, but in this case one can rewrite these equations to switch from spectral components $g_j ( x)$ to the same components but shifted by frequency, namely $g_j (x - (\omega_j - \omega_1))$. The equations take form
\begin{multline}
	\label{eq:spectral_sys}
	g_j ( x - ( \omega_j - \omega_1) ) \cdot \big[ - i x + i ( \omega_j - \omega_1) + \gamma_j \big] +
	\\
	+ \sum_i g_i^\dagger ( - x - ( \omega_i - \omega_1) )  \epsilon^*_{ji} = \frac{ i }{ 2 \omega_j } ( \vec \Pi_j \vec{ \underline{\nu}}_j ( p + x )).
\end{multline}

This set represents a system of linear algebraic equations and could be solved by applying Kramer's rule. Using the solution we obtain the expression for plus or minus quadratures of $j$-th mode $G_j^{(\pm)}$ defined as follows 
\begin{equation}
	\notag
	G_j^{(\pm)} ( x ) = \frac{ g_j (x) \pm g_j^\dagger (-x ) }{ \sqrt 2 },
\end{equation}
and calculate spectral densities of these quadratures. 

In coincidence with expectations in absence of parametric modulation both plus and minus quadratures of first mode have equal spectral densities. If modulation is applied the plus quadrature \smash{$G_1^{(+)}$} appears to be squeezed and the minus quadrature \smash{$G_1^{(-)}$} to be antisqueezed. The bigger is modulation coefficient the more significant the (anti-) squeezing effect is. It could be easily shown that squeezing is also limited by a factor of two in this case. 

\section{Readout}

The important question is whether it is possible to see this squeezing in output light. The answer to this question is positive. 

Output fluctuations $a^o$ are defined by input fluctuations $a^i$ and light inside cavity $b$, in particular for phase quadrature of $a^o$ the following expression is valid:
\begin{equation}
	\notag
	a_2^o = - a_2^i - 2 \sqrt{ \Gamma \tau } b_1.
\end{equation}
Plugging the solution obtained in previous section for $g_{1,2}$ into the equation~\eqref{eq:decomposition} we can thus express the output fluctuations through quantities $g_i$. 
\begin{figure}[t]
	\begin{center}
		\includegraphics[width = \linewidth]{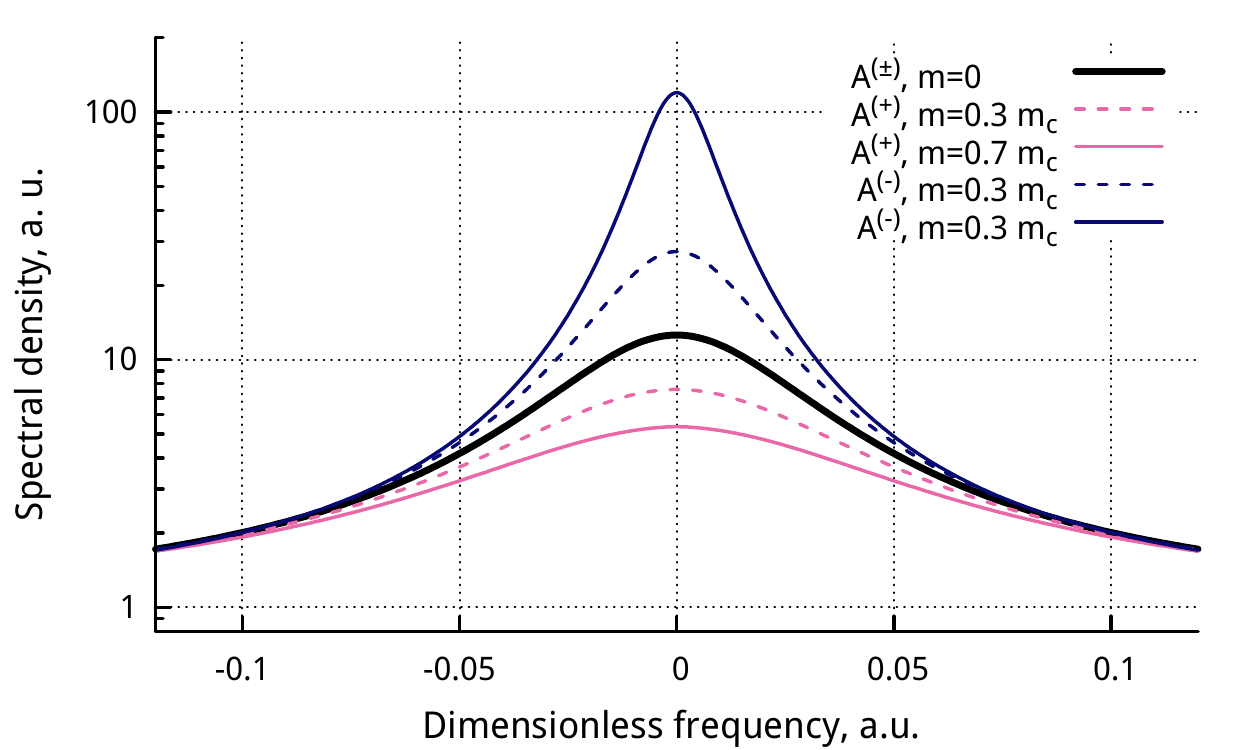}
		\caption{Spectral density of output field quadratures $A_1^{(+)}$ and $A_1^{(-)}$ (defined in eqn.~\eqref{eq:adef}) versus dimensionless frequency for different values of modulation coefficient. Bold black trace portrayes coincident plots of both $A^{(+)}$ and $A^{(-)}$ quadratures corresponding to the case of absent modulation.
		}
		\label{fig:output}
	\end{center}
\end{figure}
It is easy to show (see appendix~\ref{app:readout}) that quantity 
\begin{equation}
	\label{eq:adef}
	A^{(\pm)}_j (x) =  \frac{ a_2^o ( x + \omega_j ) \pm a_2^{o} (x - \omega_j )}{ \sqrt 2 }
\end{equation}
is proportional to the quadrature $G^{(\pm)}_j (x)$. In time domain measurement of this quantity is equivalent to multiplication of measured phase quadrature with cosine of frequency $\omega_j$ with proper phase:
\begin{equation}
	\notag
	a(t) \cos ( \omega_j t + \phi_\text{M} ) \to a (x + \omega_j) e^{ i \phi_\text{M}} + a (x - \omega_j) e^{ - i \phi_\text{M}}. 
\end{equation}
Setting $\phi_\text{M} = 0$ or $\phi_\text{M} = \frac \pi 2 $  we obtain $A_j^{(+)} $ or $ A_j^{(-)}$. 

Estimations of $A^{(\pm)}_j$ spectral density demonstrate that it is possible to see squeezing in output light. The spectral densities of these quantities are plotted in fig.~\ref{fig:output} being normalized to the shot noise level. The limitation of squeezing by factor of two shows up in output light too.

\section{Conclusion}
In this paper we have shown the possibility of quadrature noise squeezing of eigen modes amplitudes in an optomechanical system. As a model of latter we have used a Fabry-Perot cavity with movable mirror and a detuned pump. The eigen modes in this system are defined by interaction between optical and mechanical degrees of freedom. 

Detuned pump transforms the mirror which is initially a free-mass into a harmonic oscillator with spring coefficient provided by means of optical rigidity. In comparison with material rigidity peculiar to usual optomechanical devices such as microtoroids or membranes etc~\cite{Thompson2008,Verhagen2012,Eichenfield2009,Vanner2011} optical rigidity is characterized by very low noises level. The another advantage of using optical springs is the very ability of spring constant manipulation by means of pump power modulation. Comparison of impact exerted by quantum and thermal noises upon quadratures  (see Appendix~\ref{app:thermal}) reveals that both fluctuations have their affects of same order. This gives us an ability to talk about quantum noise squeezing of optomechanic quadratures in gravitational wave detectors (Advanced LIGO~\cite{Harry_2010}, Einstein Telescope~\cite{hild2011sensitivity}) and in prototypes (Glasgow University prototype~\cite{Sorazu2013}, AEI Hannover prototype~\cite{Dahl2012} and Gingin facility~\cite{Zhao2006})

Despite the fact that we consider a feedback in our model, the feedback itself is unnecessary to achieve the squeezing as it serves only to stabilize the eigen frequencies of the system. The stabilization could be performed by means of an auxiliary pump with power significantly lower than one in main pump. However implementation of feedback is more feasible in small-scale experimental setups like gravitational-wave antenna prototypes~\cite{Sorazu2013,Dahl2012,Zhao2006}. 

Realization of proposed scheme in experiment could demonstrate non-classical behavior of optomechanical mode which could be useful for tests of quantum mechanics applied to macroscopic mechanical objects. We would like to underline it is not pure optical or pure mechanical, but an {\em optomechanical} degree of freedom that may exhibit quantum behavior. Also a quantum state if created in an eigen mode could be transmitted into another one (for example, by modulation of pump with difference frequency $(\omega_2-\omega_1)$). This gives us a potential playground for creation and transmission of quantum states between optomechanical modes which could be useful for problems of quantum information.  


\acknowledgements

The authors would thank Stefan Hild for stimulating discussions. We  also  are grateful to Yanbei Chen, Farid Khalili and Stefan Danilishin. Authors are supported  by the Russian Foundation for Basic Research Grant No. 08-02-00580-a and NSF grant  PHY-0967049.


\appendix
\section{Derivation of initial equations}
\label{app:notations}

The hamiltonian of considered system (without feedback) could be written as follows~\cite{Law1995}:
\begin{equation}
	\notag
	H = \hbar \omega_c B^\dagger B + \frac{p^2 }{ 2 \mu } - \hbar K B^\dagger B y + H_\text{bath} + H_\text{pump}.
\end{equation}
Here $B$ ($B^\dagger$) is annihilation (creation) operator of optical mode, $y$ ($p$) is position (momentum) operator of movable mirror.  Also $\omega_c = \omega_0 - \Delta$ is the frequency of cavity fundamental mode closest to the pump frequency $\omega_0$. The difference between these frequencies is denoted as~$\Delta$, please note the sign of this detuning: for the pump tuned onto the right slope of optical resonance curve (blue detuned pump) $\Delta$ is \emph{positive}. $K = \omega_c/L $ is the coupling coefficient between light mode and mirror motion. $H_\text{bath}$ is the hamiltonian describing interaction of the system with bath and the bath itself. $H_\text{pump}$ is the same for pump.  

If we then suppose the optical operators to consist of strong classical part $\bar B$ (which is real positive number characterizing mean optical power $P$ inside cavity: $\bar B = \sqrt{ P/ \hbar \omega_0}$) and quantum fluctuations $b$ so that $B = \bar B + b$ and switch to the frame rotating with frequency $\omega_0$, then the linearized hamiltonian should take the form
\begin{equation}
	\notag
	H = - \hbar \Delta b^\dagger b + \frac{ p^2 }{ 2 \mu } - \hbar K \bar B y ( b^+ + b )  + H_\text{bath} + H_\text{pump}. 
\end{equation}

One then should write down Heisenberg equations for position and momentum of mirror and amplitude ($b_1$) and phase ($b_2$) operators of light defined in following way: 
\begin{equation}
	\notag
	b_1 \equiv [ b + b^+ ]/ \sqrt 2 ; \qquad b_2 \equiv - i [ b - b^+ ]/ \sqrt 2.
\end{equation}
{
To obtain the equations of motion~\eqref{eq:sys1} one should eliminate $p$ and $b_2$ and derive the forces acting from the pump and bath.  

In consideration above we have not given concrete expressions to the bath and pump hamiltonians so in order to perform inclusion of damping, feedback and optical fluctuational forces in a correct way we use semiclassical approach writing down the equations of motion in spectral form~\cite{2003Chen}:}
\begin{align*}
	\notag
	b_1 \Big[ ( \Gamma - i \Omega & )^2 + \Delta^2 \Big] + y \Big[ \sqrt{ \frac{ P \omega_0 }{ \hbar L^2 }} \Delta \Big] =
	\\
	& = - \sqrt{ \frac{ \Gamma }{ \tau }} \left[ a_1^{i} ( \Gamma - i \Omega ) + a_2^{i} \Delta \right];
	\\
	- \Omega^2 y - \frac{ 2 }{ \mu c }& \sqrt{ P \hbar \omega_0 } b_1 =  i \Omega \alpha_\text{fb} ( - a_2^{i} - 2 \sqrt{ \Gamma \tau } b_1 ).
\end{align*}

Here $\Omega$ is spectral frequency, $\tau$ is the time it takes light to travel between mirrors (one way), $\alpha_\text{fb}$ is a coefficient of feedback, i.e. the force fed to the mirror is equal to $f_\text{fb} = - \mu \alpha_\text{fb} \dot a_2^o$. The dimensionless parameters are defined as follows. 
	\begin{gather*}
		z = \sqrt{ \frac{ \mu \Delta }{ 2 \hbar \tau }} y;\quad x = \frac{ \sqrt 2 \Omega }{ \sqrt{ \Gamma^2 + \Delta^2 }}; \quad 
		g = \frac{ 2 \sqrt 2 \Gamma }{ \sqrt{ \Gamma^2 + \Delta^2 }}; 
		\\
		A = \frac{ 2}{ \Gamma^2 + \Delta^2 }\: \sqrt{ \frac{ 2 P \omega_0 \Delta}{ \mu L c} } ;
		\\
		\alpha = \frac{ 2 \alpha_\text{fb}}{ \sqrt { \Gamma^2 + \Delta^2 }} \sqrt{ \frac{ \mu \Delta \Gamma }{ \hbar }}.
	\end{gather*}
{
Switching to dimensionless parameters from dimensional ones using the definitions listed above yields the system 
\begin{align*}
	-x^2 b_1 -  i x g b_1 +& 2 b_1 + A z = \nu_1 \equiv 
	\\
	& \frac{ - g }{ \sqrt{ \Gamma \tau }} \left[ a_1^i \left( \frac g 2 - i x  \right) + a_2^i \sqrt{ 2 - \frac{ g^2 }{ 4 }} \right],
	\\
	- A b_1 + & i x \alpha b_1 - x^2 z = \nu_2 \equiv \frac{ i x \alpha}{ 2 \sqrt{ \Gamma \tau }} a_2^i. 
\end{align*}
Finally by transferring from frequency domain to time one following rule $- i \Omega \to \partial_t$ one can write down the system~\eqref{eq:sys1}
}

For numerical calculations through this paper we have used the following set of parameters 
{
\begin{equation}
	\notag
	A = 0.90; \quad g = 0.1; \quad \alpha = 0.1.
\end{equation}
}

{
\section{Thermal fluctuations}
\label{app:thermal}

In order to account thermal fluctuations in the model one should include corresponding fluctuational force in the right part of the equation of motion for mechanical degree of freedom. This results in formal replacement 
\begin{equation}
	\label{eq:thermalquantum}
	\nu_2 = \frac{ i x \alpha }{ 2 \sqrt{ \Gamma \tau }} a_2^i \to \frac{ i x \alpha }{ 2 \sqrt{ \Gamma \tau }} a_2^i - x^2 z_\text{th}.
\end{equation}
Here $- x^2 z_\text{th}$ is thermal fluctuational force. If we consider coating Brownian noise as the source of this force (in Advanced LIGO coating fluctuations are dominant thermal noises), then its spectral density is given by expression~\cite{Harry2002Thermal}:
\begin{align}
	S_{z\text{th}} &= \frac{\mu \Delta}{2\hslash \tau} \frac{2k_BT(1-\sigma^2)}{\pi^{3/2} f\, w Y} (\phi_\|+ \phi_\bot),
	\notag
	\\
	\notag
	\phi_\| &= \frac{(1+\sigma)(1-2\sigma^2)}{\sqrt \pi w Y(1-\sigma)}\left[ \frac{Y_1d_1\phi_1}{1-\sigma_1^2} + \frac{Y_2d_2\phi_2}{1-\sigma_2^2}\right],
	\\
	\phi_\bot &= \frac{Y}{\sqrt \pi w (1-\sigma^2)}\left[ \frac{(1+\sigma_1)(1-2\sigma_1)d_1\phi_1}{Y_1(1-\sigma_1)} + \right.
	\nonumber
	\\
	\notag
 	&\qquad + \left. \frac{(1+\sigma_2)(1-2\sigma_2)d_2\phi_2}{Y_2(1-\sigma_2)}\right].
\end{align}
Here $f$ is spectral frequency (measured in Hz), $w$ is a radius of a beam spot on mirror's surface, $Y,\, \sigma$ are Young modulus and Poisson ratio of substrate respectively, $Y_{1,2},\, \sigma_{1,2}$ are the same quantities for alternating layers, $d_{1,2} =N\lambda/4n_{1,2}$ are total thicknesses of quater wavelength layers, N is the number of layers' pairs, $\lambda$ --- optical wavelength, $n_{1,2}$ are refraction indices of layers. $T$ is the temperature of mirrors coating, $k_\text{B}$ is Bolzmann's constant. The numerical parameters that we use for estimations are listed in table~\ref{param}. 
\begin{table}[h!]
	\begin{center}
		\caption{Parameters used for numerical calculations of coating Brownian noise.}\label{param}
		\begin{tabular}{|c|c|c|c|}
			\hline
			~Parameter~& ~substrate & ~Ta$_2$O$_5$ layer & ~SiO$_2$ layer \rule{0mm}{5mm}\\[0.5mm]
			\hline
			$T$, K &\multicolumn{3}{c}{290}~\vline\\
			$\lambda$, m &\multicolumn{3}{c}{$1.064\times10^{-6}$}~\vline\\
			\hline
			$N$ &~ - & $20$ &~ $20$\\
			$n$ &~ 1.45 &~ 2.035 &~ 1.45\\
			$Y$, Pa &~ $72\times10^{9}$ &~ $140\times10^{9}$ &~ $72\times10^{9}$\\
			$\sigma$ &~ $0.17$ &~ $0.23$ &~ $0.17$\\
			$\phi$ &~ $4\times10^{-10}$ &~ $2\times10^{-4}$ &~ $4\times10^{-5}$\\
			\hline
		\end{tabular}
	\end{center}
\end{table}

As the thermal noise is not correlated to vacuum noises, inclusion of the last term in equation~\eqref{eq:thermalquantum} reveals in estimations of spectral densities of quadratures only by additional term in expression for spectral density of $\nu_2$:
\begin{equation}
	\notag
	S_{\nu_2} \to S_{\nu_2} + x^4 S_{z\text{th}}. 
\end{equation}

The convenient factor of thermal noise influence on eigen mode amplitude spectral density is the following ratio:
\begin{equation}
	\notag
	\xi = \sqrt{ \frac{ S_\text{th} (\omega_1 ) }{ S_\text{q} ( \omega_1) } } \Bigg|_{m = 0 }.
\end{equation}
Here $S_\text{th}$ is the spectral density of output field quadrature $A_1^{(+)}$ in case of only thermal noises present and $S_\text{q}$ is the spectral density of the same quadrature provided by quantum noises. Both spectral densities are calculated for the case of absent modulation. Obviously there is a possibility to speculate on quantum noise squeezing if factor $\xi$ is lesser than unity. 

Calculations done for parameters of Advanced LIGO~\cite{Harry_2010} yield the value $\xi_\text{aLIGO} = 0.82$ which means that despite the fact that limitations imposed on the antenna sensitivity by coating Brownian noises are smaller then quantum noise imposed ones, their influence on the eigen mode amplitude are still comparable with the influence of latter. However the estimations for future antenna Einstein Telescope~\cite{hild2011sensitivity} provide much more optimistic value $\xi_\text{ET} = 0.15$.  

Estimations for some experimental prototypes show that thermal noises are dominant in these devices. In particular for Glasgow University prototype~\cite{Sorazu2013} $\xi_\text{GP} = 2.7$, for AEI Hannover prototype~\cite{Dahl2012} $\xi_\text{AEI} = 1.7 $ and for Gingin facility~\cite{Zhao2006} $\xi_\text{Gingin} = 3.8 $. The parameters used for these estimations are listed in table~\ref{tab:prototype}. Calculations for Einstein Telescope differ from another devices by used wavelength $\lambda = 1.55\ \mu$m and temperature $T = 10$ K. 

\begin{table}[h!]
	\begin{center}
		\caption{Numerical parameters of Advanced LIGO (aLIGO), Einstein Telescope~(ET), Glasgow Prototype (GP), AEI Hannover Prototype and Gingin High Optical Power Test Facility used for numerical estimations of thermal noises influence. }\label{tab:prototype}
		\begin{tabular}{|c|c|c|c|c|c|}
			\hline
			{}  & aLIGO & ET & GP & AEI & Gingin \\ \hline
			Arm length, m & $4 \times 10^3$ & $10 \times 10^3$ & $10$ & $10$ & $77$ \\
			Reduced mass, kg & 10 & 50 & 0.1 & 0.025 & 0.4 \\
			Beam spot radius, m & 0.05 & 0.09 & 0.01 & 0.01 & 0.01 \\
			Power in arm, kW & $5$ & 18  & 10  & 10 & 40 \\
			\hline
		\end{tabular}
	\end{center}
\end{table}

}
\section{Readout}
\label{app:readout}

In output field we measure phase quadrature $a^o_2$ which is proportional to $b_1$ that carries information about squeezing. Further we discuss which combination of spectral components of $b_1$ contains squeezing evidence taking into account that corresponding combination of $a^o_2$ components represents the same combination with some input fluctuations $a_2^i$ added. 

For spectral components of $b_1$ shifted in frequency domain by amount of $\omega_j$ one can write the following expression using eqn.~\eqref{eq:decomposition}:
\begin{equation}
	\notag
	b_1 ( x + \omega_j ) = \sum_i V_i ( g_i (x - [ \omega_i - \omega_j ] ) + g_i^\dagger ( - x - [ \omega_i + \omega_j] )).
\end{equation}
Here we use freedom in definition of $\vec v_i$ and let $V_i = ( \vec v_i)_1 = ( \vec v_i )^*_1$. An equation similar to one above could be written for $b_1 ( x - \omega_j)$:
\begin{equation}
	\notag
	b_1 ( x - \omega_j ) = \sum_i V_i \big( g^\dagger_i ( - x - [ \omega_i -  \omega_j ] ) + g_i ( x - [\omega_i + \omega_j]) \big).
\end{equation}
We now take sum or difference of these quantities and obtain expression containing sum of quadratures $G_j^{(\pm)}$ that are of our interest
\begin{multline}
	\notag
	B_j^{(\pm)} \equiv \frac{ b_1 ( x + \omega_j ) \pm b_1 ( x - \omega_j)}{\sqrt  2} = 
	\\
	\sum_i \frac{ V_i }{ \sqrt 2 } \Big[ \big( g_i ( x - ( \omega_i - \omega_j)) \pm g_i^\dagger ( - x - ( \omega_i - \omega_j ) ) \big) + 
	\\
	+ \big( g_i ( x - ( \omega_i + \omega_j ) ) \pm g_i^\dagger ( - x - ( \omega_i + \omega_j )) \big) \Big] = 
	\\
	= \sum_i V_i \Big[ G_i^{(\pm)} ( x - ( \omega_i - \omega_j)) + G_i^{(\pm)} ( x - ( \omega_i + \omega_j)) \Big]. 
\end{multline}

Remind that we assumed $g_i$ to be slow amplitudes which means that spectral components of these quantities are situated close to zero frequency. Note that due to resonant multiplyers (expressions in square brackets) in eqn.~\eqref{eq:spectral_sys} one can estimate the width of band containing $g_i$ spectra with order of corresponding mode decay rate $\gamma_i$ which is much smaller than difference between eigen frequencies $\omega_i - \omega_j$ (for $i\neq j$). This consideration is also valid for quadratures $G_i$ hence in definition of \smash{$B_j^{(\pm)}$} all summands differ from zero in different frequency bands (not overlapping because of narrowness), or in other words in each frequency band there is not more than one non-zero summand in this definition. In particular at frequencies close to zero one can use quite exact expression 
\begin{equation}
	\notag
	B_j^{(\pm)} (x) = V_j G_j^{(\pm)} (x). 
\end{equation}

As we actually measure $a_2^o$ we need to consider a combination of spectral components of this quantity containing $B_j^{(\pm)}$. It is obvious that corresponding combination is given by the following expression
\begin{multline}
	\notag
	A_j^{(\pm)} (x) \equiv \frac{ a_2^o ( x + \omega_j) \pm a_2^o ( x - \omega_j) }{ \sqrt 2 } = 
	\\
	= \frac{ a_2^i ( x + \omega_j) \pm a_2^i ( x - \omega_j) }{ \sqrt 2 } + 2 \sqrt{ \Gamma \tau} \cdot B_j^{(\pm)} (x),
\end{multline}
which represents a sum of desired quadrature and input fluctuations. 




\end{document}